\documentclass[a4paper,british,aps,prb,preprint,superscriptaddress,showkeys]{revtex4-1}
\usepackage[latin9]{inputenc}
\usepackage{color}
\usepackage{babel}
\usepackage{amsmath}
\usepackage{amssymb}
\usepackage{graphicx}
\usepackage{esint}
\usepackage[unicode=true,pdfusetitle,
 bookmarks=true,bookmarksnumbered=false,bookmarksopen=false,
 breaklinks=false,pdfborder={0 0 1},backref=false,colorlinks=false]
 {hyperref}
\usepackage{breakurl}

\makeatletter

\providecommand{\tabularnewline}{\\}

\makeatother

\begin{document}

\title{Modelling a Bistable System Strongly Coupled to a Debye Bath: A Quasiclassical
Approach Based on the Generalised Langevin Equation}

\thanks{This article is dedicated to William G. Hoover on his 80th birthday}

\author{L. Stella}

\email[Corresponding author:]{l.stella@qub.ac.uk}

\affiliation{Atomistic Simulation Centre, School of Mathematics and Physics, Queen's
University Belfast, University Road, Belfast BT7 1NN, Northern Ireland,
UK}

\author{H. Ness}

\affiliation{Department of Physics, Faculty of Natural and Mathematical Sciences,
King's College London, Strand, London WC2R 2LS, UK}

\author{C.D. Lorenz}

\affiliation{Department of Physics, Faculty of Natural and Mathematical Sciences,
King's College London, Strand, London WC2R 2LS, UK}

\author{L. Kantorovich}

\affiliation{Department of Physics, Faculty of Natural and Mathematical Sciences,
King's College London, Strand, London WC2R 2LS, UK}

\begin{abstract}
Bistable systems present two degenerate metastable configurations
separated by an energy barrier. Thermal or quantum fluctuations can
promote the transition between the configurations at a rate which
depends on the dynamical properties of the local environment (\emph{i.e.},
a thermal bath). In the case of classical systems, strong system-bath
interaction has been successfully modelled by the Generalised Langevin
Equation (GLE) formalism. Here we show that the efficient GLE algorithm
introduced in Phys. Rev. B \textbf{89}, 134303 (2014) can be extended
to include some crucial aspects of the quantum fluctuations. In particular,
the expected isotopic effect is observed along with the convergence
of the quantum and classical transition rates in the strong coupling
limit. Saturation of the transition rates at low temperature is also
retrieved, in qualitative, yet not quantitative, agreement with the
analytic predictions. The discrepancies in the tunnelling regime are due to 
an incorrect sampling close to the barrier top. 
The domain of applicability of the quasiclassical GLE is also discussed.
\end{abstract}


\maketitle

\section{Introduction\label{sec:introduction}}

The Generalised Langevin Equation (GLE)\cite{Zwanzig1961,Mori1965a,Mori1965b,Zwanzig1973}
is a stochastic equation which describes a mechanical system subject
to a random force or noise. At variance with the original Langevin
equation\cite{Lemons1997} used to model Brownian motion, the
GLE can deal with a wider range of random forces, \emph{e.g.}, with
non-trivial time correlations. The GLE also provides a theoretical
framework for the definition of the frequency dependent linear response
of a mechanical system through the fluctuation-dissipation theorem.\cite{Kubo1966,Zwanzig2001}

The GLE has been employed to extend the classical transition rate
theory of Kramers\cite{Kramers1940,Haenggi1990} to include the coupling
to realistic thermal baths with a frequency dependent spectral function.\cite{Pollak1986a,Pollak1989,Tucker1991,Carmeli1982,Carmeli1983,Marchesoni1983}
It is even possible to formulate a quantum GLE\cite{Ford1965,Benguria1981,Ford1987,Ford1988a}
to model the deviation from the classical transition rate theory at
low temperature due to dissipative tunnelling (\emph{i.e.}, tunnelling without energy conservation).\cite{Pollak1986b,Pollak1986c,Ford1988b}
In fact, the problem of dissipative tunnelling has been solved
theoretically by using path-integral techniques.\cite{Caldeira1981,Caldeira1983a,Caldeira1983b,Grabert1988a}
Alternative approaches which make use of a $c$-number
\footnote{Following Dirac\cite{Dirac1926}, $c$-numbers are ``classical numbers'' which always commute (\emph{e.g.}, complex numbers),
while $q$-numbers are ``quantum numbers'' which do not commute in general (\emph{e.g.}, linear operators).}
quantum GLE\cite{Schmid1982,Eckern1990,Stockburger2002,Barik2003a,Banerjee2004,Dammak2009,Kantorovich2016}
are in principle better suited for numerical simulations since they are
based on real-time equations of motion. However, the existing approaches
are either more computationally demanding than the classical GLE or
their applicability to the strong system-bath coupling regime has
not been fully demonstrated, yet.

In this article we use a $c$-number quantum GLE approach similar
to the quasiclassical Langevin equation of Schmid\cite{Schmid1982}
or the quantum thermal bath of Dammak \emph{et al.}\cite{Dammak2009}.
On the other hand, the GLE approach considered in this article is
able to model a wider class of thermal bath and
takes full advantage of the algorithmic development introduced
in Ref.~\onlinecite{Stella2014}. In this way, we are
able to investigate the strong coupling regime of a bistable system
coupled to a Debye bath, \emph{i.e.}, a thermal bath with a sharp
frequency cut-off. Here we employ the adjective ``quasiclassical''
to distinguish this approximate scheme from the exact $c$-number quantum
GLE introduced in Ref.~\onlinecite{Kantorovich2016}. 
In particular, the main topic of this article is the domain
of applicability of the quasiclassical GLE in the case of low temperature
and strong system-bath coupling,
while a detailed discussion of the $c$-number quantum GLE formalism
can be found in Ref.~\onlinecite{Kantorovich2016}.

As in the case of similar quasiclassical approximations,\cite{Eckern1990} the
GLE approach used in this article fails to model tunnelling (with or without energy conservation)
at low temperature, while dissipative tunnelling --- especially in the weak coupling 
regime --- can be tackled by real-time GLE approaches which include quantum
corrections to the force field.\cite{Banerjee2002a,Banerjee2002b,Barik2003a}
However, the addition of these quantum corrections comes at
a computational price. In this article we demonstrate that the results of 
the quasiclassical GLE approach --- the computational cost of which is essentially 
equal to that of its classical counterpart --- are in surprisingly 
good agreement with the analytic predictions for the quantum transition 
rates\cite{Pollak1986c,Grabert1987} in the strong coupling limit. 
In particular, the isotope effect and the convergence of the quantum and classical 
transition rates in the strong coupling limit are correctly modelled.

The article is organised as follows: in Sec. \ref{sec:quasiclassical},
the quasiclassical GLE is introduced along with the relevant terminology.
In Sec. \ref{sec:bistable}, the model bistable potential is defined,
the main properties of the Debye bath discussed, and the capabilities
of the quasiclassical GLE to model the quantum probability densities
demonstrated for a light test particle (hydrogen or deuterium). In
Sec. \ref{sec:rates} the classical and quasiclassical transition rates
are investigated as a function of the particle mass and system-bath
coupling strength. Finally, in Sec. \ref{sec:discussion} and \ref{sec:conclusions}
the results of the quasiclassical GLE are discussed in detail and the
conclusions about the domain of applicability of the quasiclassical
GLE approach are drawn.

\section{Quasiclassical GLE\label{sec:quasiclassical}}

In this Section, we complete the GLE formalism introduced in 
Ref.~\onlinecite{Stella2014} to include the quantum delocalisation
at low temperature. For the sake of simplicity, we consider only the
case of one particle in one spatial dimension. The generalisation
to many particles in three spatial dimensions is straightforward.
This extension is similar to other approaches to the quantum Langevin
equation based on the quantum fluctuation-dissipation theorem (QFDT).
\cite{Ford1965,Benguria1981,Ford1987,Ford1988a,Schmid1982,Cortes1985,Banerjee2004,Dammak2009,Ceriotti2009b} 

The quasiclassical GLE is integrated by means of the following complex
Langevin equations:

\begin{equation}
\begin{aligned}\dot{r} & =\frac{p}{m}\;,\\
\dot{p} & =-\frac{\partial V\left({r}\right)}{\partial r}+\sum_{k=0}^{K}G_{k}\left(r\right)s_{1}^{\left(k\right)}\;,\\
\dot{s}_{1}^{\left(k\right)} & =-\frac{s_{1}^{\left(k\right)}}{\tau_{k}}+\omega_{k}s_{2}^{\left(k\right)}-\frac{\mu}{m}G_{k}\left(r\right)p+\sqrt{\frac{2\mu h\left(\omega_{k}\right)k_{B}T}{\tau_{k}}}{\bf \xi}_{1}^{\left(k\right)}\;,\\
\dot{s}_{2}^{\left(k\right)} & =-\frac{s_{2}^{\left(k\right)}}{\tau_{k}}-\omega_{k}s_{1}^{\left(k\right)}+\sqrt{\frac{2\mu h\left(\omega_{k}\right)k_{B}T}{\tau_{k}}}{\bf \xi}_{2}^{\left(k\right)}\;,
\end{aligned}
\label{eq:quasiclassical_eoms}
\end{equation}
where $r$ and $p$ are the physical degrees of freedom (DoFs), $s_{1}^{\left(k\right)}$
and $s_{2}^{\left(k\right)}$ are $K+1$ pairs of auxiliary DoFs,
$m$ is the physical mass, $\mu$ is the mass of the auxiliary DoFs,
$V\left(r\right)$ is the physical potential,\footnote{In fact, $V\left(r\right)$ includes a ``polaronic'' correction
due to the system-bath interaction. For the sake of simplicity, we
do not treat this correction explicitly in the rest of this article
and we treat $V\left(r\right)$ as if it was independent of the system-bath
coupling strength. } $G_{k}\left(r\right)$ are the (dimensional) coupling strengths,
$\tau_{k}$ are the relaxation times of the pair of auxiliary DoFs,
$\omega_{k}\ge0$ are the frequencies of the auxiliary DoFs, and ${\bf \xi}_{1}^{\left(k\right)}$
and ${\bf \xi}_{2}^{\left(k\right)}$ are pairs of uncorrelated sources
of white Gaussian noise, \emph{i.e.}, stochastic processes with zero
average,
\begin{equation}
\left\langle \xi_{1}^{\left(k\right)}\left(t\right)\right\rangle =\left\langle \xi_{2}^{\left(k\right)}\left(t\right)\right\rangle =\left\langle \xi_{1}^{\left(k\right)}\left(t\right)\xi_{2}^{\left(k^{\prime}\right)}\left(t^{\prime}\right)\right\rangle =\left\langle \xi_{2}^{\left(k\right)}\left(t\right)\xi_{1}^{\left(k^{\prime}\right)}\left(t^{\prime}\right)\right\rangle =0\;,
\end{equation}
and the following 2-point correlation function:\footnote{Higher order correlations function are computed by means of the Wick's
theorem.} 
\begin{equation}
\left\langle \xi_{2}^{\left(k\right)}\left(t\right)\xi_{2}^{\left(k^{\prime}\right)}\left(t^{\prime}\right)\right\rangle =\left\langle \xi_{1}^{\left(k\right)}\left(t\right)\xi_{1}^{\left(k^{\prime}\right)}\left(t^{\prime}\right)\right\rangle =\delta_{kk^{\prime}}\delta\left(t-t^{\prime}\right)\;.
\end{equation}
Following the derivation used in Ref.~\onlinecite{Stella2014},
the exact integration of the equations of motion (EoMs) of the \emph{complex}
auxiliary DoFs, $s^{\left(k\right)}=s_{1}^{\left(k\right)}+is_{2}^{\left(k\right)}$,
and its substitution into the second line of Eq. (\ref{eq:quasiclassical_eoms})
yield the quasiclassical GLE
\begin{equation}
\begin{aligned}\dot{r} & =\frac{p}{m}\;,\\
\dot{p} & =-\frac{\partial V\left(r\right)}{\partial r}-\int_{-\infty}^{t}{\rm d}t^{\prime}\;\mathcal{K}\left(t-t^{\prime};r\left(t\right),r\left(t^{\prime}\right)\right)\frac{p\left(t^{\prime}\right)}{m}+{\eta}\left(t;r\left(t\right)\right) \;,
\end{aligned}
\label{eq:gle-quantum}
\end{equation}
with the (classical) memory kernel defined as

\begin{equation}
\mathcal{K}\left(t-t^{\prime};r,r^{\prime}\right)=\mu\sum_{k=0}^{K}\left[G_{k}\left(r\right)G_{k}\left(r^{\prime}\right)e^{-\frac{1}{\tau_{k}}\left(t-t^{\prime}\right)}\cos\left(\omega_{k}\left(t-t^{\prime}\right)\right)\right]\theta\left(t-t^{\prime}\right)\;,\label{eq:kernel}
\end{equation}
and the (complex) coloured Gaussian noise
\begin{equation}
\begin{aligned}{\eta}\left(t;r\right) & =\mbox{Re}\left\{ \sum_{k=0}^{K}\sqrt{\frac{2\mu h\left(\omega_{k}\right)k_{B}T}{\tau_{k}}}\int_{-\infty}^{t}\mbox{d}t^{\prime}\;G_{k}\left(r\right)e^{-\left(\frac{1}{\tau_{k}}+i\omega_{k}\right)\left(t-t^{\prime}\right)}{\bf \xi}^{(k)}\left(t^{\prime}\right)\right\}
\end{aligned}
\;,\label{eq:noise-quantum}
\end{equation}
where ${\bf \xi}^{(k)}={\bf \xi}_{1}^{(k)}+i{\bf \xi}_{2}^{(k)}$.
Note that the noise includes the quantum weight\footnote{The quantum weight is the ratio between the internal energy of an
independent bosonic oscillator, $\frac{\hbar\omega}{2}\coth\left(\frac{\hbar\omega}{2k_{B}T}\right)$,
and the internal energy of an independent classical oscillator, $k_{B}T$,
of equal frequency, $\omega$. In both cases, ``independent'' means
``in the limit of vanishing small system-bath interaction''.}
\begin{equation}
h\left(\omega\right)=\frac{\hbar\omega}{2k_{B}T}\coth\left(\frac{\hbar\omega}{2k_{B}T}\right)=\begin{cases}
\frac{1}{2}\frac{\hbar\left|\omega\right|}{k_{B}T} & {\rm for}\quad T\to0\mbox{ or }\omega\to\infty\;,\\
1 & {\rm for}\quad T\to\infty\mbox{ or }\omega\to0\;.
\end{cases}\label{eq:quantum-fluctuations}
\end{equation}

The parameters $\tau_{k}$ and $\omega_{k}$, along with the coupling
strengths, $G_{k}\left(r\right)$, can be either deduced from the
\emph{first principle} system-bath Lagrangian (in the classical case,
\cite{Kantorovich2008,Stella2014}) or fitted to an approximate
memory kernel, $\mathcal{K}$, obtained from benchmark molecular
dynamics simulations.\cite{Ness2015,Ness2016,Gottwald2015,Gottwald2016} While the
second case is most useful in practise, the exact mapping between
the first principle Lagrangian and the parametrisation of the GLE
kernel ensures that both the equilibrium and relaxation of the physical
DoFs are correctly modelled, at least in the classical case. 

The coloured Gaussian noise defined in Eq. (\ref{eq:noise-quantum})
has zero average, $\left\langle {\eta}\left(t;r\left(t\right)\right) \right\rangle =0$,
while the 2-point correlation function is given by
\begin{equation}
\left\langle {\eta}\left(t;r\left(t\right)\right) {\eta}\left(t^{\prime};r\left(t^{\prime}\right)\right) \right\rangle =k_{B}T\mathcal{K}^{\left(q\right)}\left(t-t^{\prime};r\left(t\right),r\left(t^{\prime}\right)\right)\;,\label{eq:gfdt-quantum}
\end{equation}
where the \emph{quantum} memory kernel is defined as

\begin{equation}
\mathcal{K}^{\left(q\right)}\left(t-t^{\prime};r,r^{\prime}\right)=\mu\sum_{k=0}^{K}\left[h\left(\omega_{k}\right)G_{k}\left(r\right)G_{k}\left(r^{\prime}\right)e^{-\frac{1}{\tau_{k}}\left(t-t^{\prime}\right)}\cos\left(\omega_{k}\left(t-t^{\prime}\right)\right)\right]\theta\left(t-t^{\prime}\right)\;.\label{eq:kernel-quantum}
\end{equation}
A crucial difference between Eq. (\ref{eq:kernel}) and Eq. (\ref{eq:kernel-quantum})
is the presence of the quantum weight in the second equation, although
we have that $\mathcal{K}^{\left(q\right)}\to\mathcal{K}$ in the
limit of either $T\to\infty$ or $\omega\to0$ (\emph{i.e.}, in the
classical limit, see Eq. (\ref{eq:quantum-fluctuations})). 

In order to faithfully reproduce the quantum delocalisation close
to a minimum of the physical potential, $V\left(r\right)$, the QFDT
must hold. This is indeed the case in the limit of infinitely many auxiliary DoFs, $K\to\infty$. In
this limit, we have that $\tau_{k}\to\infty$ (see Sec. \ref{sec:bistable})
and we can rewrite the quantum kernel as\cite{Cortes1985} 
\begin{equation}
\begin{gathered}\mathcal{K}^{\left(q\right)}\left(t-t^{\prime};r,r^{\prime}\right)=\mu\sum_{k=0}^{\infty}\left[h\left(\omega_{k}\right)G_{k}\left(r\right)G_{k}\left(r^{\prime}\right)\cos\left(\omega_{k}\left(t-t^{\prime}\right)\right)\right]\theta\left(t-t^{\prime}\right)\\
=\frac{2}{\pi}\int_{-\infty}^{\infty}\frac{\mbox{d}\omega}{\omega}\;h\left(\omega\right)J\left(\omega;r,r^{\prime}\right)\cos\left(\omega\left(t-t^{\prime}\right)\right)\;,
\end{gathered}
\label{eq:kernel-quantum-J}
\end{equation}
where we have introduced the spectral density:
\begin{equation}
J\left(\omega;r,r^{\prime}\right)=\frac{\pi\mu\omega}{2}\sum_{k=0}^{\infty}G_{k}\left(r\right)G_{k}\left(r^{\prime}\right)\delta\left(\omega-\omega_{k}\right)\theta\left(\omega\right)\;.\label{eq:spectral_density_delta}
\end{equation}
By means of Eq. (\ref{eq:spectral_density_delta}) and Eq. (\ref{eq:quantum-fluctuations}),
the 2-point correlation function of the noise, Eq. (\ref{eq:gfdt-quantum}),
can be expressed as
\begin{equation}
\left\langle {\eta}\left(t;r\left(t\right)\right) {\bf \eta}\left(t^{\prime};r\left(t^{\prime}\right)\right) \right\rangle =\frac{\hbar}{\pi}\int_{-\infty}^{\infty}\mbox{d}\omega\;\coth\left(\frac{\hbar\omega}{2k_{B}T}\right)J\left(\omega;r,r^{\prime}\right)\cos\left(\omega\left(t-t^{\prime}\right)\right)\;,
\end{equation}
which is a most familiar form of the QFDT. Note that the noise
correlation saturates in the limit of $T\to0$, while in the classical
case (obtained by fixing $h\mbox{\ensuremath{\left(\omega\right)}}=1$)
we have that $\left\langle {\eta}\left(t;r\left(t\right)\right) {\eta}\left(t^{\prime};r\left(t^{\prime}\right)\right) \right\rangle \to0$.

The QFDT is only approximately satisfied for a finite number of auxiliary DoFs.
In this case, one can still define a spectral density
\begin{equation}
J\left(\omega;r,r^{\prime}\right)=\frac{\mu\omega}{2}\sum_{k=0}^{K}G_{k}\left(r\right)G_{k}\left(r^{\prime}\right)\left[\frac{\tau_{k}}{1+\left(\omega-\omega_{k}\right)^{2}\tau_{k}^{2}}+\frac{\tau_{k}}{1+\left(\omega+\omega_{k}\right)^{2}\tau_{k}^{2}}\right]\theta\left(\omega\right)\label{eq:spectral_density}
\end{equation}
so that
\begin{equation}
\mathcal{K}\left(t-t^{\prime};r,r^{\prime}\right)=\frac{2}{\pi}\int_{-\infty}^{\infty}\frac{\mbox{d}\omega}{\omega}\;J\left(\omega;r,r^{\prime}\right)\cos\left(\omega\left(t-t^{\prime}\right)\right)\;.
\end{equation}
However, Eq. (\ref{eq:kernel-quantum-J}) does not hold strictly because
of the frequency dependence of $h\left(\omega_{k}\right)$, which
cannot be factorised out of the (finite) summation over the index,
$k$, of the auxiliary degrees of freedom. In practise, numerical
convergence of the correlation functions and other figures of merit
must be verified for each model of the environment. In the case of
the Debye bath considered in Sec. \ref{sec:bistable}, convergence
is quickly achieved (namely, for $K=50$) in the weak coupling regime, 
although extra care must be paid to the strong coupling regime.\cite{Stella2014} 

\section{Model bistable system coupled to a Debye bath\label{sec:bistable}}

We model the bistable system by means of the quartic double-well potential
\begin{equation}
V\left(r\right)=V_{b}\left[1-\left(\frac{r}{r_{min}}\right)^{2}\right]^{2}\;,
\end{equation}
where $V_{b}$ is the barrier height and the two equivalent minima
are located at $r=\pm r_{min}$. To investigate the possible relevance
of the isotope effect, the mass of the test particle is taken either
as $m=m_{H}=1.0079$ amu (hydrogen) or $m=m_{D}=2.0141$ amu (deuterium).
This model is artificial, but simple enough to provide neat results
about the transition rates (see Sec. \ref{sec:rates}).
On the other hand, it can also serve as a first step towards the application of
the quasiclassical GLE to model hydrogen-bonded solids and liquids.

\begin{figure}
\includegraphics[width=8cm]{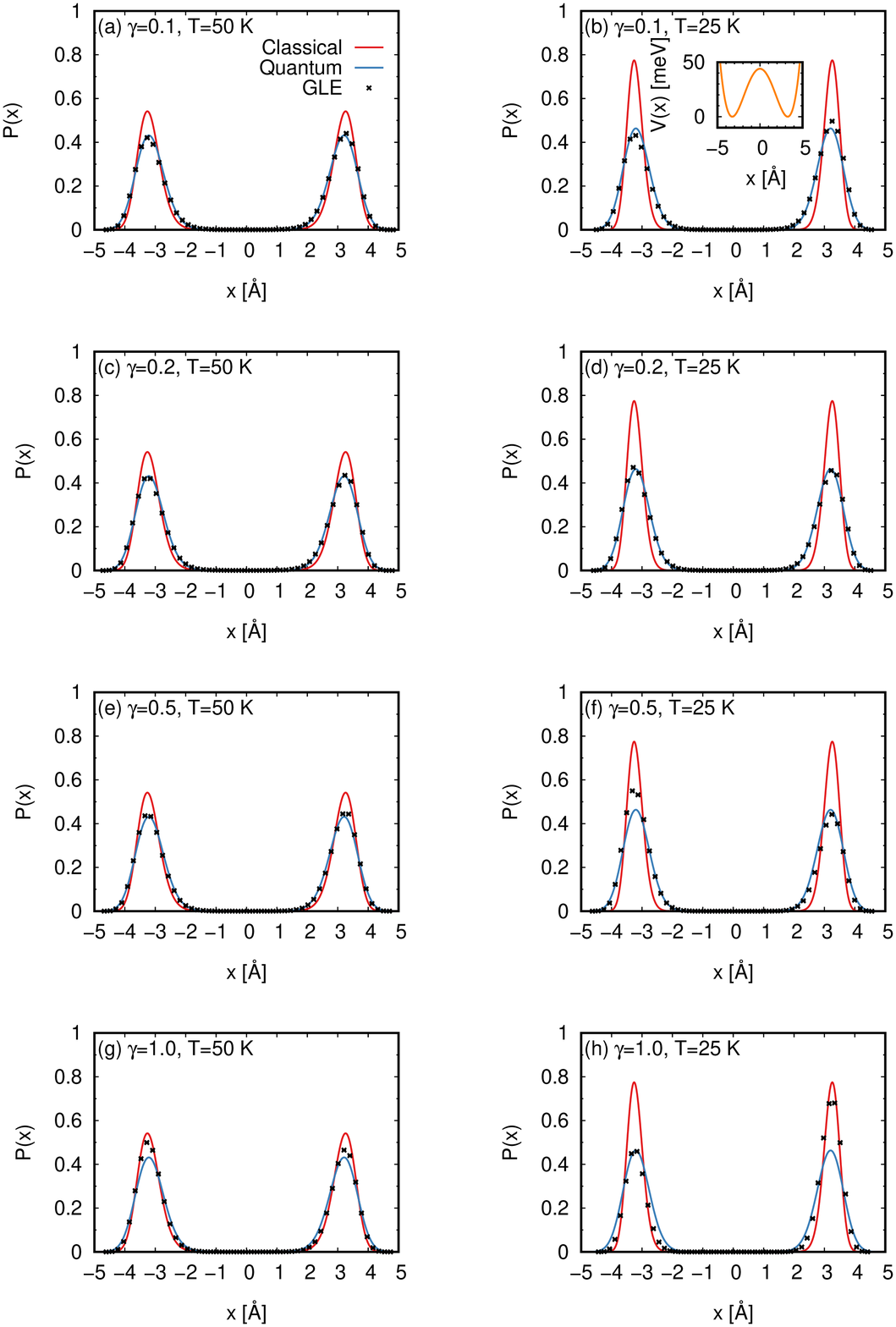}

\caption{Probability density of a hydrogen atom in a bistable potential well
(see the inset of panel (b)) for $T=50$ K (left panels) and $T=25$
K (right panels) and several values of the dimensionless coupling
strength, $\gamma$ (black points with error bars).
The probabilities and the errors have been estimated from the position histograms of
the corresponding equilibrated GLE simulations.
Analytic estimates of the classical (red lines) and quantum (blue lines) 
probability densities are also reported. 
The dynamics in the cases of $T=25$ (low temperature)
and $\gamma\ge0.5$ (strong coupling) are not ergodic (see text).\label{fig:fig1}}

\end{figure}

\begin{figure}
\includegraphics[width=8cm]{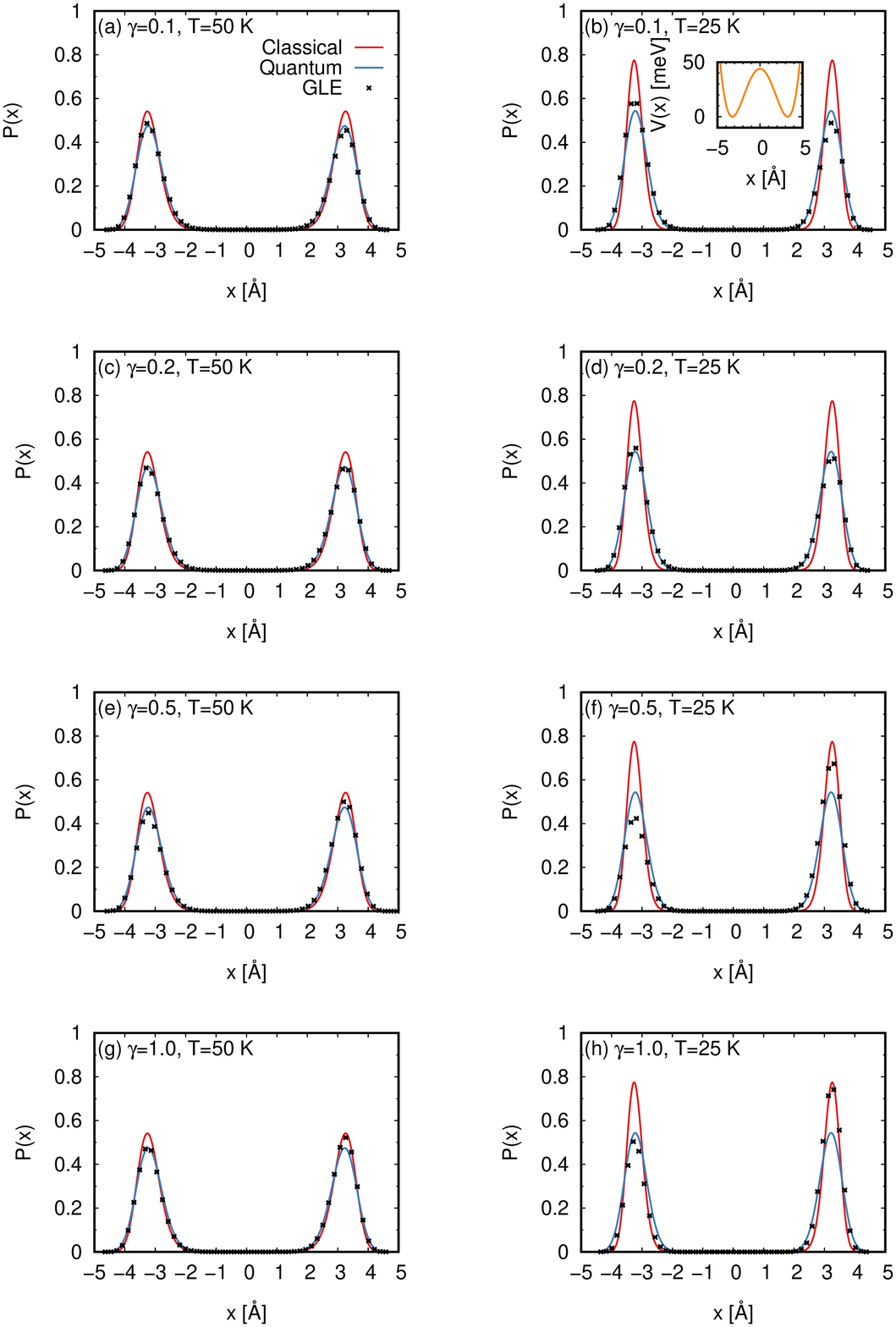}

\caption{Probability density of a deuterium atom in a bistable potential well
(see the inset of panel (b)) for $T=50$ K (left panels) and $T=25$
K (right panels) and several values of the dimensionless coupling
strength, $\gamma$ (black points with error bars).
The probabilities and the errors have been estimated from the position histograms of
the corresponding equilibrated GLE simulations.
Analytic estimates of the classical (red lines) and quantum (blue lines) 
probability densities are also reported. 
The dynamics in the cases of $T=25$ (low temperature)
and $\gamma\ge0.5$ (strong coupling) are not ergodic (see text).\label{fig:fig2}}

\end{figure}

A natural unit of energy is provided by the Debye energy of the bath,
$k_{B}T_{D}=\hbar\omega_{D}$. The barrier height is then fixed to
be $V_{b}=3k_{B}T_{D}$ and the potential minima are defined using
\begin{equation}
r_{min}=\sqrt{\frac{8V_{b}}{m\Omega_{0}^{2}}}\;,
\end{equation}
after the harmonic frequency of the two equivalent
minima has been fixed at $\Omega_{0}=0.8\left(\sqrt{m_{H}/m}\right)\omega_{D}$.
The presence of the square root of the mass ratio makes the harmonic
constant (\emph{i.e.}, the second derivative of the potential at $r=\pm r_{min}$),
$m\Omega_{0}^{2}$, a geometric parameter independent of the particle
mass, $m$, as expected. The selected values of the barrier height and harmonic constant 
make possible to sample the probability densities (see Fig. \ref{fig:fig1} and \ref{fig:fig2})
and the transition rates (see Fig. \ref{fig:fig4} and \ref{fig:fig4})
by direct molecular dynamics simulations.
In the case of the hydrogen mass, the choice of the harmonic frequency agrees 
with the example considered in Ref.~\onlinecite{Stella2014}.

The Debye bath is defined by means of its Debye temperature, $T_{D}$,
the dimensionless system-bath coupling strength, $\gamma$, and the
auxiliary mass, $\mu$. In particular, we consider the values $T_{D}=170$
K and $\gamma=0.1$, $0.2$, $0.5$, $1.0$. Having fixed, $\mu=m$,
the parameters $G_{k}$,\footnote{For the sake of simplicity, here we assume that the $G_{k}$ are trivial
(\emph{i.e.}, constant) function of $r$.} $\tau_{k}$, and $\omega_{k}$ in Eq. (\ref{eq:quasiclassical_eoms})
depends on $T_{D}$ and $\gamma$, only. Following Ref.~\onlinecite{Stella2014},
we choose a uniform sampling of the frequency interval $\left[-\omega_{D},\omega_{D}\right]$, \emph{i.e.},
$\omega_{k}=\frac{k}{K}\omega_{D}$, with $k=0,1,2,\dots,K$.
We then write that $G_{k}=g_{0}c_{k}$, where $g_{0}=\gamma\left(m/m_{H}\right)\Omega_{0}^{2}$
and
\begin{equation}
c_{k}=\begin{cases}
\frac{1}{\omega_{D}}\sqrt{\frac{3}{\left(2K+1\right)}} & \mbox{if }k=0\;,\\
\frac{1}{\omega_{D}}\sqrt{\frac{6}{\left(2K+1\right)}} & \mbox{if }k>0\;.
\end{cases}
\end{equation}
Once again, the presence of the mass ratio makes the parameters $G_{k}$
independent of the particle mass, $m$, and, in the case of the hydrogen
mass, the choice of the parameters agrees with the example considered
in Ref.~\onlinecite{Stella2014}. For the sake of simplicity, we choose an equal decay time for all the auxiliary DoFs,
\begin{equation}
\tau_{k}=\tau=\lambda\frac{\left(2K+1\right)}{2\omega_{D}}\;,
\end{equation}
with the auxiliary constant $\lambda$ defined through the self-consistent
equation
\begin{equation}
\frac{\lambda}{\pi}=\left(1+2\sum_{k=1}^{K}\frac{1}{1+k^{2}\lambda^{2}\left(1+\frac{1}{2K}\right)^{2}}\right)^{-1}\;\label{eq:width_param}
\end{equation}
in order to retrieve the exact behaviour of the memory kernel in the two limits of $\omega\to 0$ and $\omega\to \infty$.\cite{Stella2014}
This choice of the parameters $\omega_k$, $c_k$ and $\tau_k$ has been preferred to a least squares fit because it yields
a more transparent convergence to the spectral density in the limit of $K\to\infty$ (see below).

By means of the Mittag-Leffler expansion of the hyperbolic cotangent,
\begin{equation}
\coth\left(z\right)=\frac{1}{z}+2z\sum_{k=1}^{\infty}\frac{1}{z^{2}+\pi^{2}k^{2}}\;,
\end{equation}
in the limit of $K\to\infty$, we can express Eq. (\ref{eq:width_param}) as
\begin{equation}
\frac{\lambda}{\pi}=\left(1+2x^{2}\sum_{k=1}^{\infty}\frac{1}{x^{2}+k^{2}\pi^{2}}\right)^{-1}=\frac{1}{x\coth\left(x\right)}\;.
\end{equation}
where
\begin{equation}
x=\frac{\pi}{\lambda}\left(1+\frac{1}{2K}\right)^{-1}\;.
\end{equation}
Hence, the self-consistent equation can be written as
\begin{equation}
x=\mbox{arcoth}\left(\frac{\pi}{\lambda x}\right)
\end{equation}
which yields
\begin{equation}
\frac{\pi}{\lambda}=\left(1+\frac{1}{2K}\right)\mbox{arcoth}\left(1+\frac{1}{2K}\right)=\frac{1}{2}\left(1+\frac{1}{2K}\right)\ln\left(1+4K\right)\;,
\end{equation}
where we have used that
\begin{equation}
\mbox{arcoth}\left(x\right)=\frac{1}{2}\ln\left(\frac{x+1}{x-1}\right)\;.
\end{equation}
Solving the last equation for $\lambda$, we can also estimate the
asymptotic behaviour in the limit of $K\to\infty$,
\begin{equation}
\lambda\sim\frac{2\pi}{\ln\left(4K\right)}\;,
\end{equation}
which yields
\begin{equation}
\tau\sim\frac{2\pi K}{\omega_{D}\ln\left(4K\right)}
\end{equation}
and the expected limit of $\tau\to\infty$ if $K\to\infty$ (see Sec.
\ref{sec:quasiclassical}).

By means of Eq. (\ref{eq:spectral_density}), we write the spectral
density of the Debye bath as
\begin{equation}
J\left(\omega\right)=\frac{2\Gamma\omega_{D}\omega}{\pi\left(2K+1\right)}\left[\frac{\tau}{1+\omega^{2}\tau^{2}}+\sum_{k=1}^{K}\left(\frac{\tau}{1+\left(\omega-\omega_{k}\right)^{2}\tau^{2}}+\frac{\tau}{1+\left(\omega+\omega_{k}\right)^{2}\tau^{2}}\right)\right]\theta\left(\omega\right)\;,\label{eq:spectral_density_debye}
\end{equation}
where the effective friction constant, $\Gamma$, is defined by the
equation
\begin{equation}
\frac{\Gamma}{m}=\frac{3}{2}\pi\gamma^{2}\left(\frac{m}{m_{H}}\right)^{2}\left(\frac{\Omega_{0}}{\omega_{D}}\right)^{4}\omega_{D}\;.\label{eq:gamma}
\end{equation}
As usual, the presence of the mass ratio makes the parameter $\Gamma/m$
independent of the particle mass, $m$, and, in the case of the hydrogen
mass, the definition of $\Gamma$ agrees with the example considered in
Ref.~\onlinecite{Stella2014}. We also note that the integral
of the spectral density 
\begin{equation}
\int_{-\infty}^{\infty}\frac{\mbox{d}\omega}{\omega}\;J\left(\omega\right)=\Gamma\omega_{D}
\end{equation}
does not depend on the number of pairs of DoFs, $K+1$, and that the spectral
density has an algebraic asymptotic behaviour
\begin{equation}
J\left(\omega\right)\sim\frac{2\Gamma\omega_{D}\omega}{\pi}\left(\frac{\tau}{1+\omega^{2}\tau^{2}}\right)\theta\left(\omega\right)\;,
\end{equation}
in the limit of $\omega\gg\omega_{D}$. It can be also proven that, in the limit of $K\to\infty$, 
the spectral density in Eq. (\ref{eq:spectral_density_debye}) converges to the expected
\begin{equation}
J\left(\omega\right)=\Gamma\omega\chi_{[0,\omega_{D}]}\left(\omega\right)\;,
\end{equation}
where $\chi_{[0,\omega_{D}]}$ is the characteristic function of the interval $[0,\omega_{D}]$.

Despite the apparent simplicity of the Debye model, the limit $K\to\infty$
is not entirely trivial \cite{Kemeny1986}. As shown in Ref.~\onlinecite{Stella2014}, 
a persistent (\emph{i.e.}, undamped)
oscillation with a frequency \emph{larger} than the Debye frequency,
$\omega_{D}$, is observed in the strong coupling regime. A thorough
discussion of this persistent oscillation is neither brief nor pertinent
to the main topic of this article and it is then left to a future
publication.

In Fig. \ref{fig:fig1} we show the probability densities obtained
by numerical integration of the quasiclassical complex Langevin equations
introduced in Eq. (\ref{eq:quasiclassical_eoms}) with\textcolor{black}{{}
$K=50$ }for the case of the hydrogen mass. The numerical integration provides
an accurate solution of the equivalent quasiclassical GLE defined in
Eq. (\ref{eq:gle-quantum}). Details of the integration algorithm
can be found in Ref.~\onlinecite{Stella2014}.
\footnote{In fact, in this work we have preferred the ``BAOAB'' algorithm
of Leimkuhler and Matthews which is known to give better estimates
of the confrontational averages.\cite{Leimkuhler13b} This choice
implies a simple rearrangements of the split algorithm employed in
Ref.~\onlinecite{Stella2014}.} For each value of the temperature, $T$, and the dimensionless coupling
strength, $\gamma$, $50$ independent trajectories have been generated
to sample the position histograms. Each trajectory is randomly started
at rest in either the left or the right minima with equal probability.
A time step of $1$ fs and $10^{8}$ steps have been used, while the
configurations in the extended phase space $\left(r,p,s_{1}^{\left(k\right)},s_{2}^{\left(k\right)}\right)$
have been recorded every $10^{4}$ steps. In each panel, we have also
indicated the classical probability density, $P_{cl}\left(x\right)\propto\exp\left(-V\left(x\right)/k_{B}T\right)$,
and the quantum probability density, $P_{q}\left(x\right)\propto\sum_{n}\left|\phi_{n}\left(x\right)\right|^{2}\exp\left(-E_{n}/k_{B}T\right)$,
where $\phi_{n}$ and $E_{n}$ are the eigenvectors and eigenvalues
of the Hamiltonian operator $H=-\left(\hbar^{2}/2m\right)\nabla^{2}+V\left(x\right)$.

From the results shown in the different panels of Fig. \ref{fig:fig1},
we can conclude that the quasiclassical GLE is rather accurate in modelling
the quantum probability density when the temperature is not too low and the coupling
is not too strong. This conclusion agrees with previous observations.\cite{Ceriotti2009b}
The capability of a quantum GLE scheme based on the QFDT to model
the quantum delocalisation in a moderately anharmonic potential has
been exploited to improve the convergence of path-integral molecular
dynamics.\cite{Ceriotti2009b,Ceriotti2011} Discrepancies at low temperature
are due to the lack of ergodicity which follows a reduced transition
rate, $\kappa_{gle}$ (see Sec. \ref{sec:rates}). Discrepancies in
the strong coupling regime are due to the non-negligible corrections
to the quantum probability density caused by the system-bath interaction.\cite{Riseborough1985}
A detailed assessment of these corrections depends on the characterisation
of the persistent oscillation of a Debye bath (see above) and it is
therefore left to a future publication. In fact, the main conclusions about
the transition rate in the strong coupling regimes (see Sec. \ref{sec:rates})
do not depend on the detailed assessment of these corrections.

To investigate the isotope effect, in Fig. \ref{fig:fig2} we show
the probability densities obtained by numerical integration of the
quasiclassical complex Langevin equations for a deuterium atom.
The same trends with decreasing $T$ and increasing $\gamma$ are
observed, even if the quantum probability densities are more localised
and the ergodicity breaking is then more severe.

\section{Transition rates\label{sec:rates}}

The transition rate, $\kappa$, has been estimated from the decay
of the position autocorrelation function\cite{Gillan1987}\textcolor{black}{
\begin{equation}
\left\langle r\left(t\right)r\left(t^{\prime}\right)\right\rangle \sim e^{-2\kappa_{gle}\left(t-t^{\prime}\right)}
\end{equation}
}sampled at\textcolor{black}{{} $T=13$, $16$, $20$, $25$, $31$,
$40$, $50$, $63$, $79$, $100$, $126$, $159$, $200$ K} and
$\gamma=0.1$, $0.2$, $0.5$, $1.0$. The remaining simulation parameters
are the same as for the trajectories used to investigate the probability density
(See Sec. \ref{sec:bistable}).
\begin{figure}
\includegraphics[width=8cm]{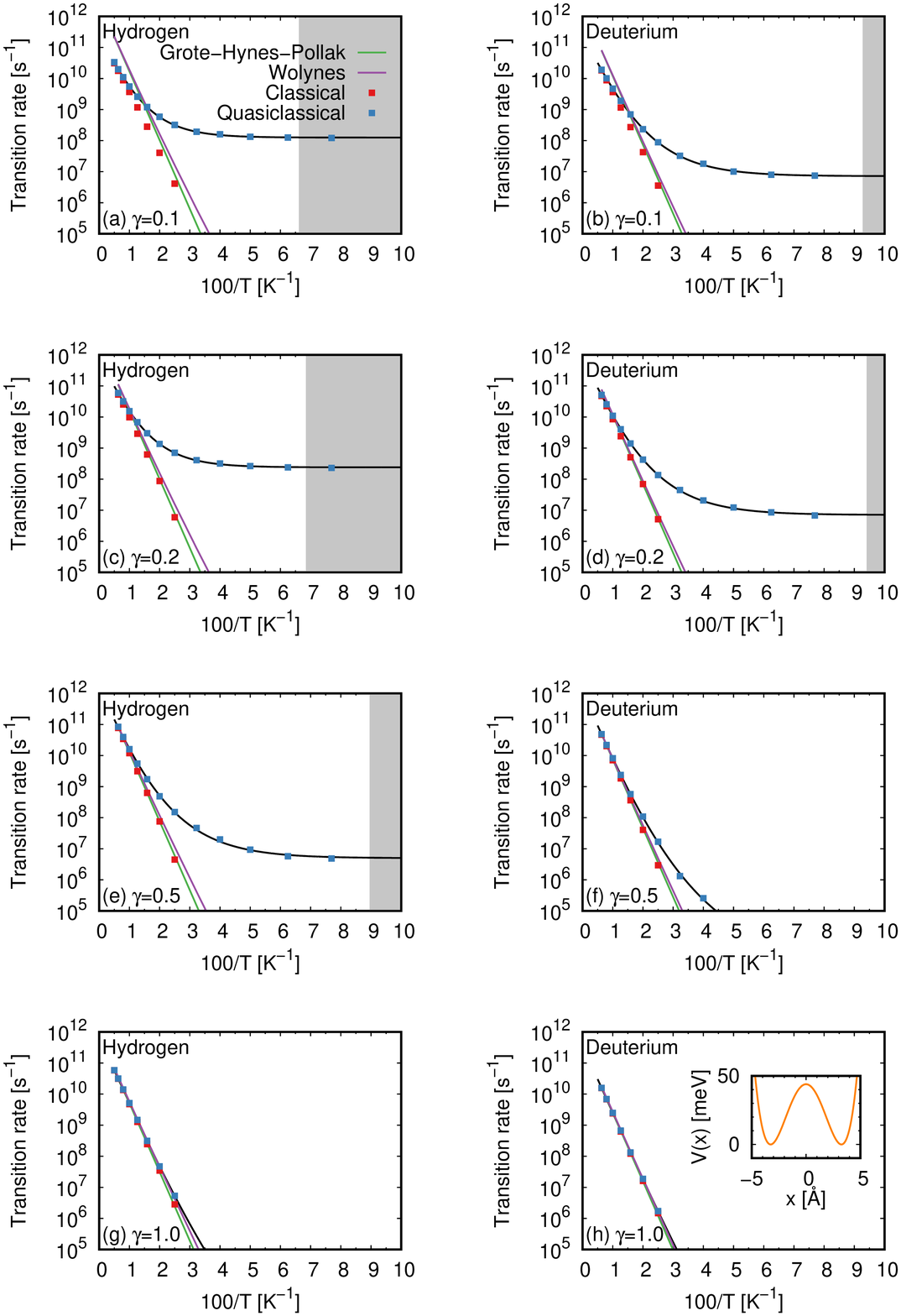}

\caption{Classical and quasiclassical GLE transition rates of a hydrogen (left
panels) or deuterium (right panels) atom in a bistable potential (see
the inset of panel (h)) as a function of the inverse temperature for
several values of the dimensionless coupling strength, $\gamma$.
For comparison, the analytic estimates from the Grote-Hynes-Pollak,
Eq. (\ref{eq:grote-hynes-pollak}), and the Wolynes, Eq. (\ref{eq:wolynes}),
formulae are also reported. The black lines are fits of the quasiclassical
GLE results (see Eq. (\ref{eq:fit})) and the regions for which $T<T_{c}$
are shaded (see Eq. (\ref{eq:Tc})). \label{fig:fig3}}
\end{figure}

In Fig. \ref{fig:fig3} we show the Arrhenius plots for the different
values of the dimensionless coupling strength in the case of the hydrogen mass
(left panels) or in the case of the deuterium mass (right panels). 
The transition rates obtained from the numerical solution of the quasiclassical GLE saturate at very
low temperature, while the transition rates from the classical GLE
(obtained by fixing $h\left(\omega\right)=1$ in Eq. (\ref{eq:quasiclassical_eoms}))
display the familiar linear behaviour.

In all the panels of Fig. \ref{fig:fig3} we also show the analytic
estimates of the classical transition rate (Grote-Hynes-Pollak\cite{Grote1980,Pollak1986a})
\begin{equation}
\kappa_{ghp}\left(T\right)=\left(\frac{\omega_{ghp}}{\omega_{b}}\right)\kappa_{tst}\left(T\right)\;,\label{eq:grote-hynes-pollak}
\end{equation}
where $-i\omega_b$ is the imaginary frequency of the barrier top (\emph{i.e.},
$-m\omega_{b}^{2}$ is the second derivative of the potential
at the barrier top) and 
\begin{equation}
\kappa_{tst}\left(T\right)=\left(\frac{\Omega_{0}}{2\pi}\right)e^{-\frac{V_{b}}{k_{B}T}}
\end{equation}
is the bare estimate of the transition state theory.\cite{Eyring1935}
In the case of a Debye bath, the Grote-Hynes-Pollak imaginary frequency, $i\omega_{ghp}$,
is given by the positive solution of the equation
\begin{equation}
\omega_{b}^{2}-\omega_{gph}^{2}\left[1+\frac{2\Gamma}{\pi m\omega_{ghp}}\arctan\left(\frac{\omega_{D}}{\omega_{ghp}}\right)\right]=0\;.\label{eq:ghp-equation}
\end{equation}
The numerical values of $\omega_{ghp}$ are reported in Table \ref{tab:table1},
for the case of the hydrogen mass, or in Table \ref{tab:table2} in the case of the deuterium mass.

The quantum transition rate (Wolynes\cite{Wolynes1981,Pollak1986b})
is approximated as\cite{Bell1959}
\begin{equation}
\kappa_{w}\left(T\right)\approx\left(\frac{\hbar\omega_{ghp}}{2k_{B}T}\right)\frac{\kappa_{tst}\left(T\right)}{\sin\left(\frac{\hbar\omega_{ghp}}{2k_{B}T}\right)}\approx\left[1+\frac{1}{24}\left(\frac{\hbar\omega_{ghp}}{k_{B}T}\right)^{2}\right]\kappa_{tst}\left(T\right)\;.\label{eq:wolynes}
\end{equation}

The agreement between the Grote-Hynes-Pollak predictions and the
classical GLE is very good in all cases, except the very weak, \emph{i.e.},
$\gamma=0.1$, coupling regime. A disagreement is expected in this
case since the transitions are limited more by energy diffusion than
spatial diffusion (the Kramers turnover problem).\cite{Haenggi1990}
Corrections to Eq. (\ref{eq:grote-hynes-pollak}) are known,\cite{Melnikov1986,Grabert1988b,Pollak1989}
but are not relevant in the strong coupling regime.

The quasiclassical GLE rates are in clear quantitative disagreement
with the Wolynes predictions in all cases, except the very strong,
\emph{i.e.}, $\gamma=1.0$, coupling regime. In particular, the quasiclassical
GLE rates saturate at a temperature well above the so-called critical
temperature,
\begin{equation}\label{eq:Tc}
T_{c}=\frac{\hbar\omega_{ghp}}{2\pi k_{B}}\;,
\end{equation}
at which Eq. (\ref{eq:wolynes}) displays an (apparent) divergence.\cite{Grabert1987,Haenggi1988,Haenggi1991}

Following Miller,\cite{Miller1977} one can use the
critical temperature, $T_{c}$, to characterise the tunnelling
through the barrier top, approximated as
$V\left(r\right)\approx V_{b}-\frac{1}{2}m\omega_{b}^{2}r^{2}$.
In this case, the tunnelling probability is
\begin{equation}
P_{t}\left(E\right)=\frac{1}{1+\exp\left(\frac{2\pi\left(V_{b}-E\right)}{\hbar\omega_{b}}\right)}=\frac{1}{1+\exp\left(\frac{V_{b}-E}{k_{B}T_{c}}\right)}\;,\label{eq:tunnelling}
\end{equation}
where $E$ is the total energy of the system. 
By means of Eq. (\ref{eq:tunnelling}), the quantum transition rate is estimated as
\begin{equation}
\kappa_{m}\left(T\right)=\left(\frac{\Omega_{0}}{2\pi K_{B}T}\right)\int_{-\infty}^{\infty}\mbox{d}E\;e^{-\frac{E}{k_{B}T}}P_{t}\left(E\right)=\left(\frac{\hbar\omega_{b}}{2k_{B}T}\right)\frac{\kappa_{tst}\left(T\right)}{\sin\left(\frac{\hbar\omega_{b}}{2k_{B}T}\right)}\;.\label{eq:miller}
\end{equation}
Note that Eq. (\ref{eq:wolynes}) and Eq. (\ref{eq:miller}) differ
only in the choice of imaginary frequency of the barrier top.
In fact, we have that $\omega_{ghp}\to\omega_{b}$ in the limit of
$\gamma\to0$ (see Eq. (\ref{eq:ghp-equation}) and Eq. (\ref{eq:gamma})).

At low temperature, $k_B T \ll V_b$, assuming that the zero-point energy is
negligible, $ \hbar \Omega_0/2\ll V_b$, one can substitute $E\approx0$ into 
Eq. (\ref{eq:tunnelling}) to approximate
$P_{t}\left(0\right)\approx e^{-\frac{V_b}{k_{B}T_c}}$. As a consequence,
tunnelling is expected to be the dominant transition mechanism for $T<T_c$. 
The numerical values of $T_{c}$ are reported in Table \ref{tab:table1}
or Table \ref{tab:table2}. The critical temperature, $T_{c}$, is
a function of both the dimensionless coupling strength, $\gamma$,
and the mass, $m$, through its dependence on $\omega_{ghp}$. The
regions corresponding to $T<T_{c}$ have been shaded in the panels
of Fig. \ref{fig:fig3}.

To help interpret the quasiclassical GLE results, we model the quasiclassical
transition rate by means of the function
\begin{equation}
\kappa_{fit}\left(T\right)=\left(\frac{A}{2\pi}\right)\exp\left(-\frac{B}{h\left(\omega^{\ddagger}\right)k_{B}T}\right)\;,\label{eq:fit}
\end{equation}
where $A$, $B$, and $\omega^{\ddagger}$ are adjustable parameters,
the values of which are reported in Table \ref{tab:table1},
\begin{table}
\begin{tabular}{|c|c|c|c|c|c|c|}
\hline 
$\gamma$ & $A$ {[}fs$^{-1}${]} & $B$ {[}eV{]} & $\omega^{\ddagger}$ {[}fs$^{-1}${]} & $T_{q}$ {[}K{]} & $T_{c}$ {[}K{]} & $\omega_{ghp}$ {[}fs$^{-1}${]}\tabularnewline
\hline 
\hline 
$0.1$ & $0.00157$ & $0.0359$ & $0.0144$ & $67.1$ & $15.1$ & $0.0124$\tabularnewline
\hline 
$0.2$ & $0.00503$ & $0.0371$ & $0.0139$ & $62.9$ & $14.6$ & $0.0120$\tabularnewline
\hline 
$0.5$ & $0.00916$ & $0.0403$ & $0.0097$ & $40.5$ & $11.2$ & $0.0092$\tabularnewline
\hline 
$1.0$ & $0.00439$ & $0.0426$ & $0.0040$ & $15.9$ & $4.6$ & $0.0038$\tabularnewline
\hline 
\end{tabular}

\caption{Numerical values of the parameters, $A$, $B$, and $\omega^{\ddagger}$
(errors on the last digit) appearing in Eq. (\ref{eq:fit}) as a
function of the dimensionless coupling strength, $\gamma$, along
with the estimates of the quantum temperature, $T_{q}$, the critical
temperature, $T_{c}$, and the Grote-Hynes-Pollak frequency, $\omega_{ghp}$,
in the case of the hydrogen mass.\label{tab:table1}}
\end{table}
or Table \ref{tab:table2}.
\footnote{One can set $B=V_{b}$ to reduce the number of adjustable parameters,
but the fits generally get worse. The difference between $B$ and
$V_{b}$ is due to the quantum fluctuations.\cite{Pollak1986c}} 
\begin{table}
\begin{tabular}{|c|c|c|c|c|c|c|}
\hline 
$\gamma$ & $A$ {[}fs$^{-1}${]} & $B$ {[}eV{]} & $\omega^{\ddagger}$ {[}fs$^{-1}${]} & $T_{q}$ {[}K{]} & $T_{c}$ {[}K{]} & $\omega_{ghp}$ {[}fs$^{-1}${]}\tabularnewline
\hline 
\hline 
$0.1$ & $0.00157$ & $0.0360$ & $0.0105$ & $48.8$ & $10.6$ & $0.0087$\tabularnewline
\hline 
$0.2$ & $0.00502$ & $0.0386$ & $0.0101$ & $43.8$ & $10.0$ & $0.0083$\tabularnewline
\hline 
$0.5$ & $0.00699$ & $0.0431$ & $0.0064$ & $25.0$ & $6.6$ & $0.0055$\tabularnewline
\hline 
$1.0$ & $0.00236$ & $0.0434$ & $0.0029$ & $11.1$ & $2.3$ & $0.0019$\tabularnewline
\hline 
\end{tabular}

\caption{Numerical values of the parameters, $A$, $B$, and $\omega^{\ddagger}$
(errors on the last digit) appearing in Eq. (\ref{eq:fit}) as a
function of the dimensionless coupling strength, $\gamma$, along
with the estimates of the quantum temperature, $T_{q}$, the critical
temperature, $T_{c}$, and the Grote-Hynes-Pollak frequency, $\omega_{ghp}$,
in the case of the deuterium mass.\label{tab:table2}}
\end{table}
\begin{figure}
\includegraphics[width=8cm]{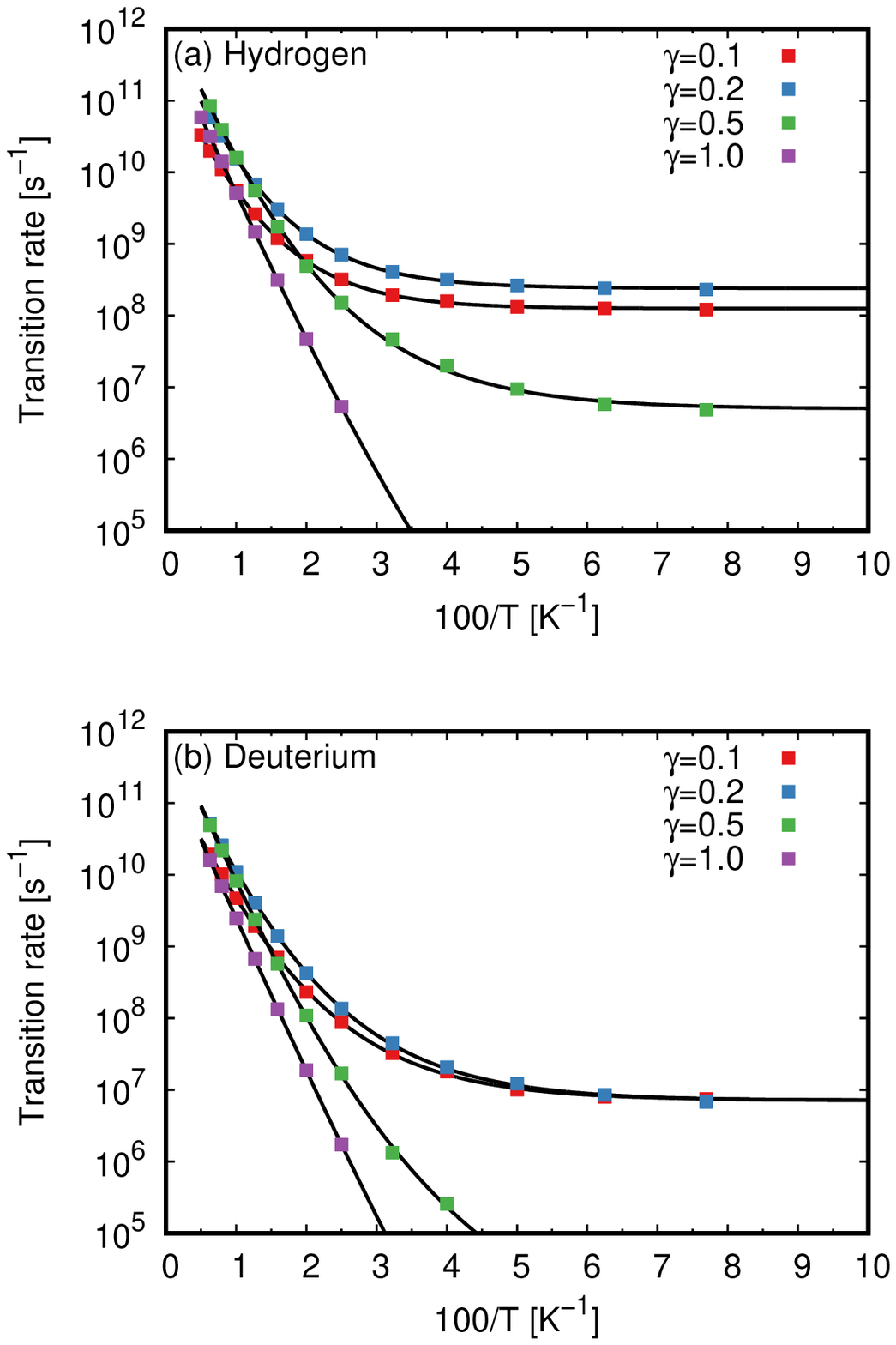}

\caption{Quasiclassical GLE transition rates of a hydrogen (panel (a)) or deuterium
(panel (b)) atom in a bistable potential as a function of the inverse
temperature (Arrhenius plots) for several values of the dimensionless
coupling strength, $\gamma$. The black lines are fits of the quasiclassical
GLE results (see Eq. (\ref{eq:fit})).\label{fig:fig4}}
\end{figure}
The global accuracy of these fits can be better appreciated from
Fig. \ref{fig:fig4}, where we have reported only the quasiclassical
GLE rates. By interpreting the exponential in Eq. (\ref{eq:fit}) as a Boltzmann factor,
we can define an effective quantum temperature, $T_{q}$, so that
\begin{equation}
\frac{V_{b}}{k_{B}T_{q}}=\frac{B}{\frac{\hbar\omega^{\ddagger}}{2}}\;.
\end{equation}
The numerical values of $T_{q}$ are reported in Table \ref{tab:table1},
for the case of the hydrogen mass, or Table \ref{tab:table2} in the case of the deuterium mass.
The quantum temperature can be used to assess the validity of
discrepancy between the quasiclassical GLE and the analytic predictions
(see Sec. \ref{sec:discussion}).

\section{Discussion\label{sec:discussion}}

The quasiclassical GLE formalism considered in this article does not
include tunnelling.\cite{Eckern1990}
It is then not surprising that it fails to model the transition rates in the deep 
quantum regime. On the other hand, it is not immediately clear why such a
formalism overestimates instead of underestimating --- as naively suggested
by the absence of tunnelling --- the quantum transition rates. In
this Section we attempt an answer by discussing in more detail the
results of Sec. \ref{sec:bistable} and \ref{sec:rates}.

First of all, from the results shown in Figs. \ref{fig:fig1} and \ref{fig:fig2},
we know that the quasiclassical GLE reproduces rather accurately the
probability density, at least close to the potential minima. In the
limit of $T\to0$ and $\gamma\to0$, the quantum fluctuations close
to the minima are entirely due to the zero-point motion which can
be characterised by the effective temperature $T_{zp}=\hbar\Omega_{0}/2k_{B}$.
In particular, we have that $T_{zp}=136$ K in the case of the hydrogen mass
or $T_{zp}=96.2$ K in the case of the deuterium mass. Those temperatures
are much larger than $T_{q}$ and $T_{c}$ in both cases. Given the
good agreement between the classical GLE rates and the Grote-Hynes-Pollak
formula for moderate to strong system-bath coupling, we can also exclude
a large contribution from the finite height of the barrier (the condition
$k_{B}T_{zp}\ll V_{b}$ is satisfied). It is then plausible that the
discrepancies between the quasiclassical GLE and the analytic predictions
originate from an incorrect sampling of the region close to the barrier
top.

The quasiclassical GLE considered in this article provides an inherently
thermal (\emph{i.e.}, classical) description of the random forces,
although the physical temperature is weighted by a correcting factor,
$h\left(\omega\right)$, which depends on the frequency of the oscillations,
$\omega$, to mimic the quantum fluctuations. In practise, the quantum
temperature, $T_{q}$, can be used to estimate the effective temperature
close to the barrier top. 
\begin{figure}
\includegraphics[width=8cm]{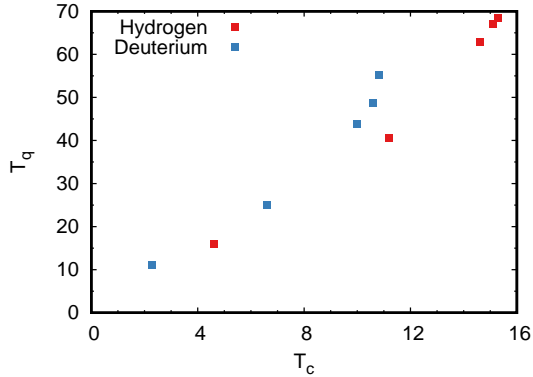}

\caption{Correlation between the effective quantum temperature, $T_q$,
and the critical temperature, $T_c$.\label{fig:fig5}}
\end{figure}

Interestingly, we observe that the ratio between $T_{q}$ and $T_{c}$
is a relatively constant function of $\gamma$ and $m$ (between $0.4$
and $0.5$, see Fig. \ref{fig:fig5}). Our results are in agreement with the
findings of Eckern \emph{et al.} 
(see Ref. \onlinecite{Eckern1990}, in particular at the end of Sec. 2.3).
This observation suggests that, despite the quantitative discrepancy
between the quasiclassical GLE and the analytic predictions, the functional
dependence of $\kappa_{gle}\left(T\right)$ on both $\gamma$ and
$m$ is qualitatively correct. In particular, the isotope effect and
the convergence of the quantum and classical transition rates in the
strong coupling limit are in good qualitative agreement with the analytic
predictions.\cite{Pollak1986c,Grabert1987} 

\section{Conclusions\label{sec:conclusions}}

In this article, we have completed the GLE formalism introduced in
Ref.~\onlinecite{Stella2014} to include the quantum delocalisation
at very low temperature. Our results confirm the applicability of
this formalism to model the equilibrium properties (\emph{e.g.}, the
probability density) of a bistable system coupled to a Debye bath.
In particular, the quasiclassical GLE formalism equally applies to both the weak
and strong coupling regimes. 

The quantitative discrepancy between the quasiclassical GLE and the
analytic predictions for the quantum transition rates has been rationalised
as the consequence of an incorrect sampling close to the barrier top.
In particular, the quasiclassical GLE predicts a saturation of the transition rate
at an effective quantum temperature, $T_q$, which is roughly twice 
as large as the expected critical temperature, $T_c$. 
Since the value of $T_c$ depends on the imaginary frequency of the barrier top (see Eq. (\ref{eq:Tc})),
we can conclude that the quasiclassical GLE effectively samples a different imaginary frequency.
On the other hand, the quasiclassical GLE accurately samples the quantum probability distribution
(at least for weak system-bath interaction) close to the minima of the potential wells.
Since the ratio between $T_q$ and $T_c$ is roughly constant,
the functional dependence of $\kappa_{gle}\left(T\right)$
on both the system-bath coupling strength, $\gamma$, and the particle
mass, $m$, is also qualitatively correct. This qualitative agreement includes the isotope effect
and the convergence of the quantum and classical transition rates
in the strong coupling limit.

Our results shed more light on the domain of applicability of a \emph{real-time} 
GLE approach to model the relaxation of \emph{quantum} dissipative
system. The simple quasiclassical approach considered in this article
ignores both the quantum corrections to the force field
\cite{Banerjee2002a,Barik2003a,Barik2003b,Banerjee2004}
and the proper treatment of the quantum fluctuations by means of the
path integral formalism\cite{Ceriotti2009b,Ceriotti2011,Richardson2009,Hele2015}.
However, it is surprisingly accurate in the limit of strong system-bath interaction,
at a computational cost essentially equal to that of its classical
counterpart.

\section*{Acknowledgements}

The authors acknowledge the financial support from EPSRC, grant EP/J019259/1.
LS is also grateful to Professors Roberto D'Agosta and Ian Ford for
many useful conversations. 

\bibliographystyle{apsrev4-1}

\end{document}